\documentclass[useAMS,usenatbib]{mn2e}

\usepackage{psfig}

\def\gs{\mathrel{\raise0.35ex\hbox{$\scriptstyle >$}\kern-0.6em
\lower0.40ex\hbox{{$\scriptstyle \sim$}}}}
\def\ls{\mathrel{\raise0.35ex\hbox{$\scriptstyle <$}\kern-0.6em
\lower0.40ex\hbox{{$\scriptstyle \sim$}}}}

\title[Multiply-imaged submm galaxy in a $z\sim 2.5$ group]{A multiply-imaged, submillimetre-selected ULIRG in a galaxy group at $z\sim 2.5$}

\author[Kneib et al.]{Jean-Paul Kneib,$^{\! 1,2}$
Paul P.\ van der Werf,$^{\! 3}$
Kirsten Kraiberg Knudsen,$^{\! 3}$
\newauthor
Ian Smail,$^{\! 4}$
Andrew Blain,$^{\! 1}$
Dave Frayer,$^{\! 5}$
Vicki Barnard$^{6,7}$
\& Rob Ivison$^{8}$\\
$^{1}$Caltech-Astronomy, MC105-24, Pasadena, CA 91125, USA\\
$^{2}$Observatoire Midi-Pyr\'en\'ees, CNRS-UMR5572,
14 Avenue E.\,Belin, 31400 Toulouse, France\\
$^{3}$Leiden Observatory, P.O.\ Box 9513, NL -- 2300 RA Leiden, 
	The Netherlands\\
$^{4}$Institute for Computational Cosmology, University of Durham,
South Road, Durham DH1 3LE, UK\\
$^{5}$Caltech-SIRTF Science Center, MC220-06, Pasadena, CA 91125, USA\\
$^{6}$Cavendish Astrophysics, University of Cambridge, Cambridge CB3 0HE\\
$^{7}$Joint Astronomy Centre, 660 N. A'ohoku Place, Hilo, HI 96720, USA\\
$^{8}$Astronomy Technology Centre, Blackford Hill, Edinburgh EH9 3HJ
}

\begin{document}

\date{Received 2003 -- --; Accepted: 2004 -- --}
\pagerange{\pageref{firstpage}--\pageref{lastpage}} \pubyear{2004}

\maketitle

\label{firstpage}

\begin{abstract}
We present observations of a remarkable submillimetre-selected galaxy,
SMM\,J16359+6612.  This distant galaxy lies behind the core of a
massive cluster of galaxies, A\,2218, and is gravitationally lensed by
the foreground cluster into three discrete images which were identified
in deep submillimetre maps of the cluster core at both 450 and
850$\mu$m.  Subsequent follow-up using deep optical and near-infrared
images identify a faint counterpart to each of the three images, with
similar red optical--near-infrared colours and {\it HST} morphologies.
By exploiting a detailed mass model for the cluster lens we estimate
that the combined images of this galaxy are magnified by a factor of
$\sim$45, implying that this galaxy would have un-lensed magnitudes
$K_s=22.9$ and $I=26.1$, and an un-lensed 850-$\mu$um flux density of
only 0.8\,mJy.  Moreover, the highly constrained lens model predicted
the redshift of SMM\,J16359+6612 to be $z=2.6\pm0.4$. We confirm this
estimate using deep optical and near-infrared Keck spectroscopy, 
measuring a redshift of $z=2.516$.  SMM\,J16359+6612 is the faintest
submm-selected galaxy so far identified with a precise redshift.
Thanks to the large gravitational magnification of this source,
we identify three sub-components in this submm galaxy, 
which are also seen in the NIRSPEC data, arguing for either a strong
dust (lane) absorption or a merger.
Interestingly, there are two other highly-amplified galaxies at almost
identical redshifts in this field (although neither is a strong submm
emitter). The three galaxies lie within a $\sim 100$\,kpc region on the
background sky, suggesting this submm galaxy is located in a dense
high-redshift group.
\end{abstract}


\section{Introduction}

Recent submillimetre (submm) surveys show that the majority of the
submm background at wavelengths of $\sim 850 \mu$m arises from a
population of distant, highly luminous infrared galaxies with
850\,$\mu$m flux densities of $\gs 1$\,mJy (Blain et al.\ 1999; Cowie
et al.\ 2002; Smail et al.\ 2002).  The steep slope of the submm counts
indicates that the integrated background is dominated by $\sim 1$\,mJy
galaxies, below the $\sim 2$\,mJy confusion limit of the deepest
blank-field 850$\mu$m surveys carried out to date (Hughes et al.\
1998), and well below the 3--8\,mJy flux density limits (3--4$\sigma$)
typically achieved in wider-field 850\,$\mu$m surveys (e.g.\ Eales et
al.\ 1999; Scott et al.\ 2002; Webb et al.\ 2003).

Submm galaxies in the 1-mJy regime can, however, only be studied at present
by exploiting a natural telescope -- a massive gravitational lens
formed by a rich clusters of galaxies -- which provide the opportunity
of overcoming both the confusion limit (by spatially magnifying an area
of the sky behind the lens) and the sensitivity limit (by gravitational
magnification of the sources within this area).  This strategy has
successfully been used by Smail, Ivison \& Blain (1997), Chapman et
al.\ (2002), Cowie et al.\ (2002) and Smail et al.\ (2002) to detect
submm sources amplified by factors typically of
1.5--4$\times$.  Unfortunately, the intrinsically faintest sources are
usually only detected at modest significance, complicating the
identification and analysis of their counterparts in other wavebands.

To investigate the properties of the mJy-population in detail we
require a high-signal-to-noise detection of a sub-mJy submm galaxy.
Such systems can arise in rare configurations where the source is
located within a small region on the background sky defined by the
caustic of the lens, which then forms multiple, highly magnified images
of the background galaxy (see also Borys et al.\ 2004).  This situation
not only provides a tool for investigating the properties of
intrinsically faint galaxies in detail, but also provides a unique
opportunity to measure the redshift of the background galaxy from the
geometry of the images combined with an accurate model of the cluster
lens, using a ray-tracing technique to triangulate the
three-dimensional position of the galaxy in the volume behind the lens.
This technique, which was demonstrated successfully by Kneib et al.\
(1996) and confirmed by Ebbels et al.\ (1998), circumvents the need for
bright optical or infrared counterparts to a submm galaxy prior to
measuring its redshift. We may thus avoid a bias in redshift
measurements towards unrepresentative, low redshift or less obscured
systems (Smail et al.\ 2002; Webb et al.\ 2003).

This paper describes the discovery of a multiply-imaged submm galaxy,
SMM\,J16359+6612, which is gravitationally lensed by the core of the
rich cluster A\,2218, and appears as three distinct sources in our
850\,$\mu$m discovery map.  We identify faint counterparts to all
three sources in deep optical and near-infrared (NIR) imaging,
consistent with their identification as three images of a single
background galaxy. We estimate its redshift 
using a detailed mass model for the cluster lens. We then confirm
the accuracy of the redshift with spectroscopic
observations and finally discuss the intrinsic properties of this system.
Throughout we will assume an $\Omega=0.3$, $\Lambda=0.7$
cosmology with $H_0=70$\,km\,s$^{-1}$\,Mpc$^{-1}$. At a redshift of
$z=2.516$ the angular scale is thus 8.06 kpc/arcsec.

\section{Observations and Results}

\subsection{Submillimetre imaging}

%
%
\begin{figure*}
\centerline{\large see JPEG file 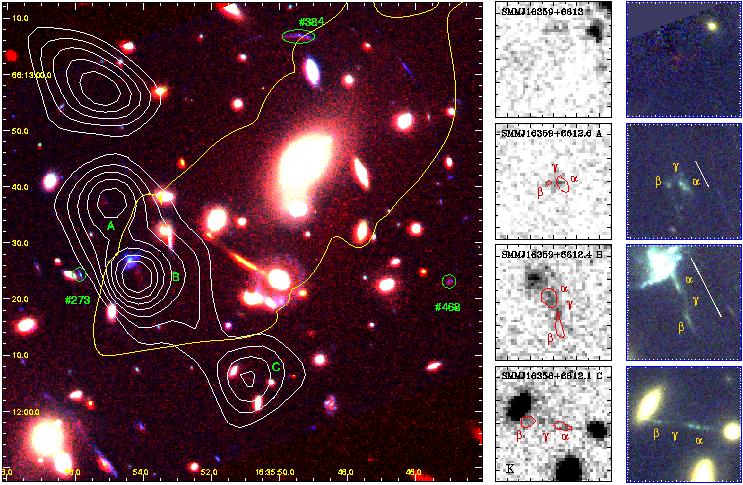}
\caption{(Left) A true-colour image of the core of A\,2218 composed
from the {\it HST} F450W (blue), {\it HST} F814W (green) and WHT/INGRID
$K_s$ (red) images.  The 850\,$\mu$m submm image from SCUBA is
overlayed as white contours at flux densities of 2.5, 3.3, 5.0, 6.6,
8.3, 10\,mJy\,beam$^{-1}$. The three images of the multiply-imaged
submm galaxy are annotated as A, B and C. We also identify the 2 other
galaxies at $z\sim 2.5$, namely the single-image \#273 and the 
fold-image \#384 and its counter-image \#468.
The yellow line shows the critical line at
$z=2.515$.
(Right) Panel of
10\arcsec$\times$10\arcsec images showing the INGRID $K_s$-band (left
column) and {\it HST} true colour image from F450W/F606W/F814W (right
column) of the four submm sources in the core of A\,2218.  The
resolution of these frames is $\sim 0.75''$ and $\sim 0.15''$
respectively. The K-band panels are centered on the submm source,
and the colour HST panels are centered on the identified optical counterpart.
The contours on the $K_s$ frames show the morphologies of the galaxy in the
F814W passband at the resolution of the $K_s$-band frame.  Note how
each of the submm sources, SMM\,J16359+6612.6, SMM\,J16359+6612.4 and
SMM\,J16358+6612.1, comprises a NIR source ($\gamma$) which is
bracketed by two features in the F814W image ($\alpha$ and $\beta$).
The morphological and photometric similarity of these three objects
suggests that they are all images of the same background source.
For SMM-A and SMM-B we have indicated in their {\it HST} true colour image
the extent of the H$\alpha$ line observed in the NIRSPEC data (white segments).
Note that there is a very faint red counterpart for the SMM\,J16359+6613
in the F814W image that does not appear in the K-band data,
and that all the bright galaxies near SMM-B and SMM-C are spectroscopically
confirmed cluster members (SMM-B is next to a star forming galaxy, and SMM-C
is next to three elliptical galaxies).
In all the images North is up and East is to the left.
\label{fig.truecol}}
\end{figure*}

We used the Submillimetre Common User Bolometer Array (Holland et al.\
1999) on the 15-m James Clerk Maxwell Telescope (JCMT) in March 1998 and 
in August 2000 to
image a $2.3'$ diameter field centred on the core of the $z=0.17$
cluster A\,2218. Total time on source excluding overheads 
was 95\,ks at a wavelength of
850\,$\mu$m with chop-throws of 45 and 90$''$ at a variety of
position angles. During the best observing conditions data were also
collected at 450\,$\mu$m (72\,ks in total without overheads).  
Zenith opacities varied
from 0.15 to 0.4 at 850\,$\mu$m and from 0.6 to 1.0 at 450\,$\mu$m.
Pointing was monitored and corrected every hour, and was found to be
accurate to better than $3''$.  The data were reduced using the
standard {\sc surf} package (Jenness \& Lightfoot 1998).  Data
reduction steps consisted of flat-fielding the demodulated data, opacity
correction, despiking, flagging bad integrations, pointing correction,
sky subtraction using the least noisy bolometers, and rebinning onto a
$1''$ grid, weighting the bolometers with the inverse variances of
their signals. Flux calibration was achieved using maps of Uranus, and
absolute flux calibration should be accurate to approximately 10\% at
850\,$\mu$m and 20\% at 450\,$\mu$m, as derived from the consistency of
the calibration measurements. The resulting 850\,$\mu$m image was
convolved with a $5''$ Gaussian to remove high-frequency noise and
subsequently deconvolved using a {\sc clean} algorithm (Hogbom 1974),
in order to remove the negative sidelobes resulting from the small chop
throw that was used for these data. After restoration with a $15''$
full width at half maximum (FWHM) Gaussian beam the resulting image has
an r.m.s.\ noise of approximately 1.5\,mJy\,beam$^{-1}$.

%
%
\begin{table*}
\caption{Properties of SMM\,J16359+6612.
All optical and NIR magnitudes are measured in a 2\arcsec diameter
apertures (Smail et al.\ 2001). The position relative to the centre of
the cD galaxy (at $16^h35^m49.21^s$, $66^d12^m44.7^s$, J2000) was
computed from the $K_s$-band data and has a relative uncertainty of
$\pm 0.5''$.  None of these sources is detected in the {\it ISO} image
on A2218 with a 3-$\sigma$ upper limit of 0.11\,$\mu$Jy (Metcalfe et
al.\ 2003). Flux density measurements and magnitudes in this table have
not been corrected for gravitational magnification.
\label{tab.results} }
\begin{tabular}{crrcccccrrrr}

ID & $\Delta\alpha$ & $\Delta\delta$ & $B-I$ & $V-I$ & $I-K$ & $J-K_s$ 
  & $K_s$ &  $f_{450}$ & $f_{850}$ & $f_{450}/f_{850}$ & Amp \\
   & (\arcsec) & (\arcsec) &   &   &   &    &
	&  mJy & mJy &  &  \\
\hline
A & 35.6 & $-$7.3 & $1.10 \pm 0.05$  & $0.45 \pm 0.05$
  &  $3.00 \pm 0.1$  & $1.2 \pm 0.2$  & $20.2 \pm 0.1$ 
  & $45\pm 9$ & $11\pm 1$ & $4.0\pm0.9$ & $14\pm 2$ \\
B & 29.6 & $-$21.1 & $1.07 \pm 0.05$ & $0.41 \pm 0.05$
  &  $3.67 \pm 0.1$ & $2.7 \pm 0.2$ & $19.5 \pm 0.1$ 
  & $75\pm 15$ & $17\pm 2$& $4.4\pm1.0$ & $22\pm 2$ \\
C & 10.2 & $-$39.1 & $1.37 \pm 0.15$ & $0.30 \pm 0.15$
  & $3.25 \pm 0.1$ & $2.0 \pm 0.2$ & $20.5 \pm 0.1$ 
  & 32$\pm 6$ & $9\pm 1$ & $3.6\pm0.8$ & $9\pm 2$ \\
\#468 & -25.7 &  $-$21.8 & $1.25 \pm 0.05$ & $0.53 \pm 0.05$
  & $2.82 \pm 0.1$ & $1.4 \pm 0.1$ & $20.0 \pm 0.1$ 
  & --- & --- & --- & $6\pm 1$ \\
\#273 & 40.3 & $-$20.7 & $0.85 \pm 0.05$ & $0.28 \pm 0.05$
  & $0.75 \pm 0.1$ & $0.0 \pm 0.1$ & $20.9 \pm 0.1$ 
  & --- & --- & --- & $30\pm 5$ \\
\hline
\end{tabular}
\end{table*}

\subsection{Optical and Near-infrared imaging}

The optical images come from the {\it Hubble Space Telescope} ({\it
HST}) WFPC2 Early Release Observations made after the SM-3a servicing
mission in January 2000. The dataset consists of 6~orbits in both F450W
(11.0\,ks) and F814W (12.0\,ks) and 5 orbits in F606W (10.0\,ks). The
reduction of these images is described in Smail et al.\ (2001).  The
final WFPC2 frames have a FWHM of $0.17''$ and 3-$\sigma$ point source
sensitivities of $B_{450}=$28.8, $V_{606}=$29.0 and $I_{814}=$28.1 in
the Vega-based system from Holtzman et al.\ (1995).

The NIR imaging used in our analysis was obtained with the INGRID
camera (Packham et al.\ 2003) on the 4.2-m William Herschel Telescope
(WHT).  The observations were obtained during INGRID commissioning on
March 22 and 23, 2000.  The dataset consists of a total of 9.1\,ks
integration in $K_s$ and 8.3\,ks in $J$-band obtained under photometric
conditions in 0.6--0.8$''$ seeing. More details on the reduction of
these data is described by Smail et al.\ (2001).  The absolute
calibration is accurate to approximately 0.03\,mag.  The final coadded
frames have 3-$\sigma$ point source sensitivities, within the seeing
disk, of $K_s=$22.0 and $J=$23.7, and seeing of $0.75''$.

\subsection{Source Identification}

We overlay the 850\,$\mu$m image on a true colour representation of the
cluster core in \ref{fig.truecol}.  Both the 850 and 450$\mu$m SCUBA
images show a number of bright submm sources located in the saddle
region in the cluster core, formed by the two brightest cluster
members. At 850\,$\mu$m, at least four sources are readily
identified above the 5-$\sigma$ detection limit of $\sim 8$\,mJy.  From
North to South these are SMM\,J16359+6613 ($\alpha= 16^h35^m55.625^s$,
$\delta= +66^d12^m59.5^s$), SMM\,J16359+6612.6, SMM\,J16359+6612.4 and
SMM\,J16358+6612.1 (Table~1). The latter three sources also appear in
the 450\,$\mu$m map, but the northern source is not detected at
450\,$\mu$m.  Significantly, the 450/850\,$\mu$m flux ratios for the
three southern sources are identical within the errors, $4.0\pm 0.4$.
In contrast, the northern source has a 3-$\sigma$ limit of
$S_{450}/S_{850}\leq 1.6$.

%
%
\begin{figure*}
\centerline{\psfig{file=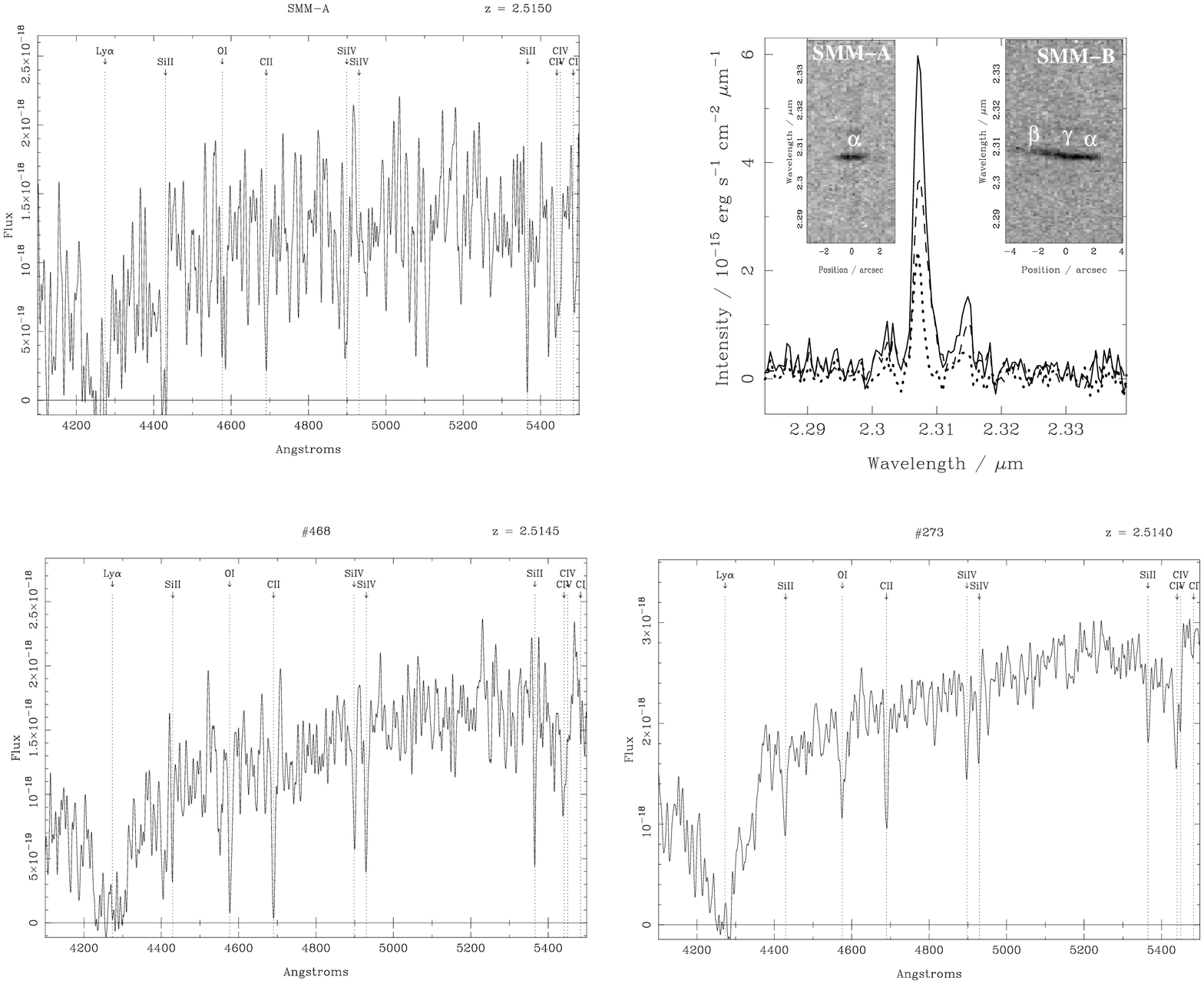,width=7in}}
\caption{ Spectroscopic identification of the three $z\sim
2.515$ galaxies behind A\,2218 from our blue-arm LRIS spectra: 
({\em Top-Left}) SMM-A, ({\em Bottom-Left}) \#468 and 
({\em Bottom-Right}) \#273.  Note the similar UV fluxes of
the three galaxies, even though they have very different submm
luminosities, and by implication very different total star formation
rates.  The three spectra have been rebinned to the same spectral
resolution of 2\AA\ and the wavelength scale is in the observed
frame. The flux is given in units of erg s$^{-1}$ cm$^{-2}$
\AA$^{-1}$. 
{\em Top-Right:} Keck/NIRSPEC $K$-band spectrum of SMM-A (dotted),
SMM-B (dashed) and
the sum of SMM-A and B (solid) showing the redshifted H$\alpha$ and
N{\sc ii} emission lines.  The inset shows the 2-d region of the
spectrum around the H$\alpha$ line in SMM-A and SMM-B demonstrating the
extended spatial morphology of the emission line in both images and the 
presence of a gravitationally amplified velocity gradient within SMM-B.
e annotated the H$\alpha$ line with the position of the different
sub component of the galaxy ($\alpha$, $\beta$ and $\gamma$).
\hfill
\label{fig.spectra}
}
\end{figure*}

Inspecting the positions of the four submm sources in the optical and
NIR data, we identify unusually red objects close to the position of
all three southern submm sources, while the field of the northern submm
source, SMM\,J16359+6613, shows a faint red counterpart visible in the
{\it HST/F814W} frames but undetected in the ground-based $K_s$-band image.
We show the relevant regions of the INGRID $K_s$ and {\it HST} frames
in Figure~\ref{fig.truecol}. It is worth noting that for all three
southern sources the relative astrometric offsets between the
$K_s$-band counterpart and the 850\,$\mu$m peak is less than 2\arcsec.

We label the optical/NIR counterparts to the three southern submm
sources from North to South: SMM-A (SMM\,J16359+6612.6), SMM-B
(SMM\,J16359+6612.4), and SMM-C (SMM\,J16358+6612.1) and list the
photometry for these within matched apertures from our optical and NIR
frames in Table~\ref{tab.results}. The colours of the three
counterparts are similar, although crowding in the various fields and
the effects of differential amplification on our fixed-aperture
photometry may explain the lack of precise agreement (the colours of
SMM-A should give the best indication of the true spectral energy
distribution of the galaxy as it is not contaminated by any 
foreground galaxy).  Nevertheless, all three counterparts show
a similar pattern of a pronounced red core surround by two blue
regions, for example in SMM-B the centre of the galaxy ($\gamma$) is
much brighter in $K_s$ than in $I_{814}$, while the southern (and
northern) extremities, $\alpha$ and $\beta$, are considerably brighter
in $I_{814}$ (Figure~\ref{fig.truecol}). A similar pattern of a red
centre bracketed by bluer regions is seen in SMM-A and C.  We
illustrate that this effect is not a result of the different spatial
resolutions of the $K_s$ and $I_{814}$ frames by contouring the
seeing-matched $I_{814}$ over the $K_s$-band frame in
Figure~\ref{fig.truecol}.

The photometric similarity of the three southern submm sources,
combined with the fact that their counterparts exhibit strong spectral
and morphological similarities, suggests that we are seeing three
images of a single luminous submm galaxy (which we will call
SMM\,J16359+6612 in the following), lying behind the cluster. The
configuration, relative brightnesses and parities of the three images
are consistent with the predictions of a detailed mass model of the
cluster core (Kneib et al.\ 1996). With the assumption that these submm
sources are three multiple images of a unique galaxy, we can predict a
redshift of $z=2.6\pm 0.4$ using the lens model of Kneib et al.\ (1996)
updated to include a recently discovered $z=5.56$ lensed galaxy (Ellis
et al.\ 2001).  For comparison, the 450/850$\mu$m flux ratio for this
galaxy corresponds to a redshift/dust temperature value
$(1+z)/(T/40{\rm K}) = 3.2 \pm 0.3$ (Blain et al.\ 2002) or a redshift
range of $z=1.9$--4.5 from the spectral energy distributions models in
Hughes et al.\ (1998).

\subsection{Keck spectroscopy}

On June 30 and July 1st 2003, we conducted deep multi-slit spectroscopy
with the Low Resolution Imaging Spectrograph (LRIS; Oke et al.\ 1995),
of sources lying in the field of the rich cluster A\,2218 (see Kneib
et al.\ 2004 for a complete description of these data and a detailled 
discussion of the mass model).  The two nights
had reasonable seeing, $\sim 0.8\arcsec$, but were not fully
photometric (with some cirrus), nevertheless we obtained a crude flux
calibration of our observations using Feige 67 and 110 as
spectrophotometric standard stars.

We observed the SMM-A source for a total of 10.6\,ks using the 600/4000
blue grism and the 400/8500 red gratings offering respectively a
spectral dispersion of 0.63\AA\ pixel$^{-1}$ in the blue and 1.86\AA\
pixel$^{-1}$ in the red.  We included in the same mask a slitlet on the
galaxy \#468 (the counter image of the $z=2.515$ fold arc \#384: Ebbels et
al.\ (1996), see also Le Borgne et al.\ (1992)).  The spectra of SMM-A
and \#468 both show strong Ly-$\alpha$ absorption and numerous metal
lines (Figure~\ref{fig.spectra}).  The derived redshift for SMM-A is
$2.515\pm 0.001$ based on the strongest interstellar absorption lines:
O{\sc i}$\lambda$1302.17\AA\ and C{\sc ii}$\lambda$1334.5\AA\ as well
as C{\sc iv}$\lambda$1548.19 and 1550.77 in absorption.
These observations also yielded a redshift for \#468 of $2.5145\pm
0.001$.  

In addition, a 9.2-ks exposure using the 400/3400 blue grism
on a second mask targeting galaxies in this field yielded a redshift
for another $z\sim 2.5$ galaxy.  This galaxy is identified as \#273 (Le
Borgne et al.\ 1992) and was misidentified as a $z=0.800$ galaxy by
Ebbels et al.\ (1998) who confused the C{\sc iii}]$\lambda$1908.7\AA\
emission line with [O{\sc ii}]$\lambda$3727\AA.  As for SMM-A and
\#384/\#468, \#273 shows prominent Ly-$\alpha$ absorption and many
metal absorption lines from which we derived a redshift of $2.514\pm
0.001$ (Figure~2).  All three galaxies are thus within 100\,km\,s$^{-1}$ of each
other in the rest frame (ignoring uncertainties due to possible
velocity offsets in the UV lines we use). Figure~\ref{fig.spectra},
shows the UV spectra of all three galaxies with the identified absorption
lines.

The counterparts of the submm sources SMM-A and SMM-B were also
independently targeted in the $K$-band using the NIR spectrograph
NIRSPEC on Keck II (McLean et al.\ 1998).  These observations were
obtained in 2002 August 18 in non-photometric conditions and
0.6\arcsec\ seeing. Eight exposures of 600\,s each were obtained, nodding
along the 0.76\arcsec$\times$42\arcsec\ slit by 8$''$ between each
exposure. Sources A and B were on the slit simultaneously.  Wavelength
calibration was achieved from the sky lines, using observations of a
bright star to correct for geometrical distortion. An average sky
spectrum was created by scaling and combining the individual spectra,
and this was then subtracted from each science observation before
co-adding and extracting the spectra. Approximate flux calibration and
telluric absorption correction was achieved using an observation of
UKIRT standard stars from the previous night.  No attempt was made to
correct for slit losses.

The flux-calibrated spectrum (Figure~\ref{fig.spectra}) reveals three
emission lines identified as H$\alpha$ and the two weaker flanking [N{\sc
ii}] lines. The H$\alpha$ line extends over 2.2\arcsec\ for SMM-A and
5.1\arcsec\ for SMM-B, which compares well with the optical
size of $\alpha$ in SMM-A and the total extent of $\alpha, \gamma$ and $\beta$
for SMM-B as shown in the {\sl HST} panels in Figure~\ref{fig.truecol} and the
two insets in Figure~\ref{fig.spectra}.
There are hints in SMM-A for a more compact distribution of
[N{\sc ii}] emission. The H$\alpha$ line is centered on 2.30755$\mu$m
placing the submm galaxy at a redshift of $z=2.5165\pm 0.0015$,
indicating that the UV interstellar lines used to derive a redshift in
the optical are blueshifted by $\sim 100$\,km\,s$^{-1}$ with respect to
the systemic velocity of the galaxy.  The restframe width of the
H$\alpha$ line corresponds to $\sim 280\pm 60$\,km\,s$^{-1}$, and there
is a hint of coherent velocity structure in the spectrum of SMM-B, with
a velocity shift of about $220\pm60$\,km\,s$^{-1}$ over an angular
distance of 2.6\arcsec\ ($\sim 1.5$\,kpc in the source plane).  The
fact that we do not see such structure in the SMM-A image is due
to the fact that only component $\alpha$ (and probably part of $\gamma$)
was in the NIRSPEC slit. Hence, this velocity shift could argue
that the complex optical/NIR morphology of this source is best explained
by a strong dust absorption in a rotating disk structure or better
by a merger.

The total H$\alpha$ line flux for the combined spectra of SMM-A and B
is $f_{H\alpha}=2.7 \times 10^{-20}$\,W\,m$^{-2}$, with a ratio of 2.3
between their respective H$\alpha$ flux densities. This corresponds to
a H$\alpha$ luminosity $L_{{\rm H}\alpha}=3.7 \times
10^8$\,L$_\odot$, or a (unlensed) star-formation rate of
11\,M$_\odot$\,yr$^{-1}$.  The narrow width of the H$\alpha$ emission
line and the observed ratio of N{\sc ii}/H$\alpha$ ($0.3\pm 0.1$)
suggest that the emission is dominated by star formation, with little
evidence for a strong AGN.

\section{Results and Discussion}
\label{sec.discussion}

The occurrence of a submm galaxy falling within the caustic lines of a
massive foreground gravitational lens provides unique constraints on
the properties of this faint submm galaxy.  Using the cluster mass
model we estimate that the background galaxy is gravitationally
amplified by a factor of $\sim$45 (integrated across all three images),
indicating that the intrinsic 850\,$\mu$m flux density of this galaxy
would be 0.8\,mJy in the absence of gravitational magnification, while
in the optical and NIR the galaxy would have magnitudes $I_{814}=26.1$
and $K_s=22.9$.  The galaxy therefore represents a serendipitously
positioned example of the submm galaxy population at flux levels of
$\sim 1$\,mJy, the population which produces the bulk of the
submm background (Blain et al.\ 1999).  This provides a unique
opportunity to compare the properties of this low-luminosity submm
galaxy with those of more luminous, submm galaxies studied in brighter,
blank-field surveys (e.g.\ Ivison et al.\ 2002; Chapman et al.\ 2003).

A second, multiply-imaged submm galaxy has been identified by Borys et
al.\ (2004), giving two high-magnification examples from the 24 cluster
lenses at $z>0.1$ mapped with sufficient sensitivity using SCUBA 
(Chapman et al.\ 2002; Smail et al.\ 2002; Knudsen et al.\ in
prep.).  The surface density of $\geq 1$\,mJy submm galaxies is around
3\,arcmin$^{-2}$ (Smail et al.\ 2002; Cowie et al.\ 2002), suggesting
that roughly ten clusters need to be surveyed to detect a
strongly-lensed mJy-flux submm galaxy.  This is consistent with the
expectation based on models for the cluster lenses of uniform
$\sim$800\,km\,s$^{-1}$ potential wells at $z\sim 0.2$, each with a critical
curve encircling 0.05\,arcmin$^{-2}$, and background submm galaxies at
$z\sim 2.4$ (see Kraiberg et al 2004 for a thorough discussion).

In our assumed cosmology the far-infrared luminosity of
SMM\,J16359+6612 is $1.0\times10^{12}$\,L$_\odot$, roughly 1.5$\times$
fainter than Arp\,220, close to the border-line between luminous
infrared galaxies and ultraluminous infrared galaxies (ULIRGs). The
star formation rate derived from the far-infrared luminosity is about
500\,M$_\odot$\,yr$^{-1}$.  Figure~3 compares the restframe
(lens-corrected) SED of the SMM\,J16359+6612 to the SEDs of Arp\,220,
the well-studied $z=2.8$ submm galaxy SMM\,J02399$-$0136 and the
$z=1.44$ Extremely Red Object and submm galaxy HR10 (Dey et al.\ 1999).  
The L$_{FIR}$/L$_{opt}$ ratio is higher than that seen for Arp\,220,
suggesting it is more obscured, and similar to the high-redshift, but
much more luminous, SMM\,J02399$-$0136 (Ivison et al.\ 1998).

%
%
\smallskip
\centerline{\psfig{file=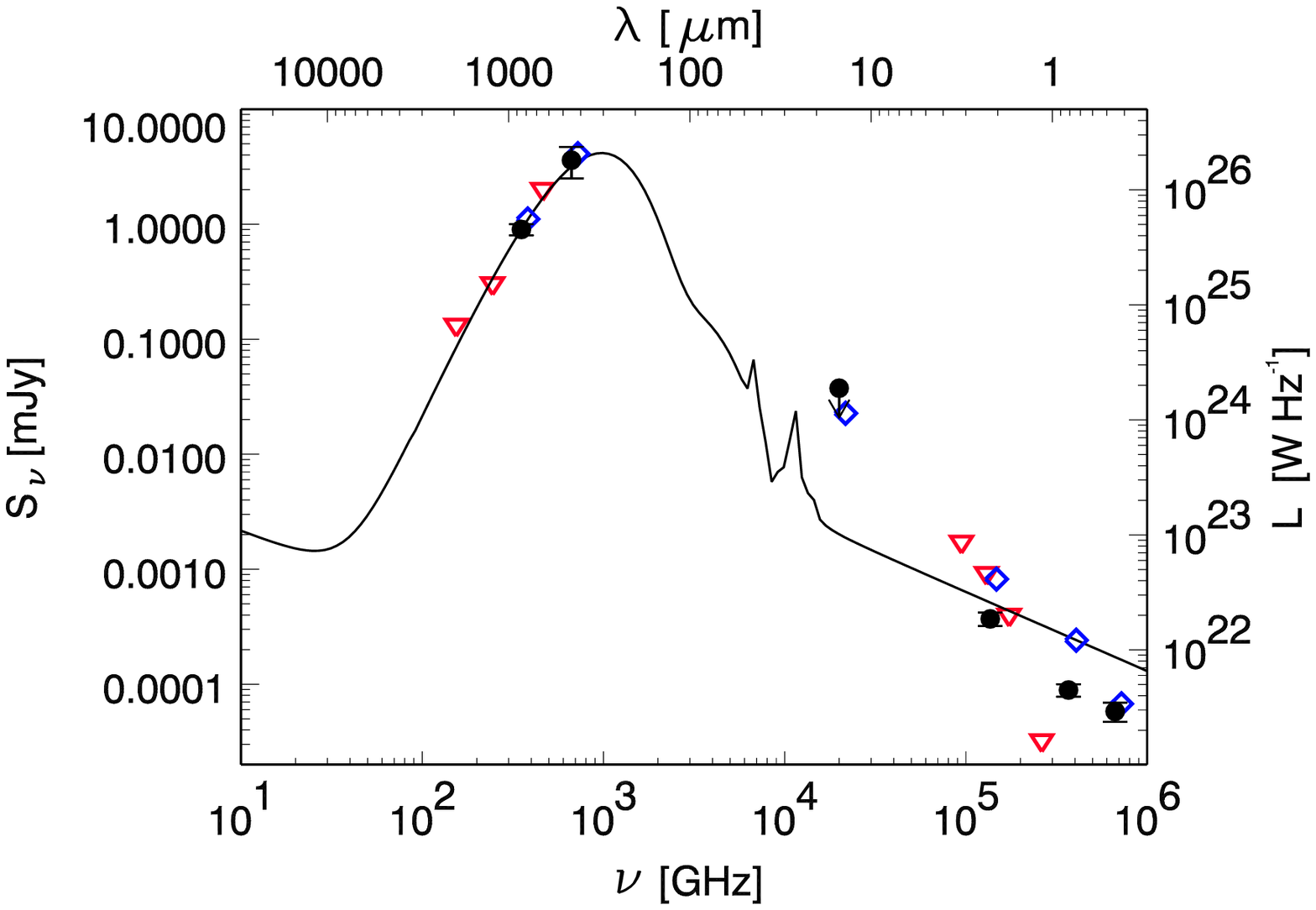,width=3.4in}}
{\small\addtolength{\baselineskip}{-1pt}
\noindent{\sc Fig.~4.} --- The observed Spectral Energy Distribution
(SED) of the multiply-imaged submm galaxy SMMJ\,16359+6612 (solid
points with error bars) compared to Arp\,220 (solid line, composite
spectrum from Anantharamaiah et al.\ 2000, Klaas et al.\ 1997,
Lisenfeld et al.\ 1996 and Surace et al.\ 2000) which is 1.5$\times$
more luminous in the far-infrared, as well as the well-studied submm
galaxy SMMJ\,02399$-$0136 (open diamonds, at $z=2.8$, Ivison et al.\ 1998)
which is 10$\times$ more luminous in the far-infrared, and the 
Extremely Red object HR10 (open triangles, at $z=1.44$, Dey et al.\ 1999)
which is 7$\times$ more luminous in the far-infrared.
All SEDs are
redshifted to match that of SMM\,J16359+6612 and scaled so they have a
similar far-infrared luminosity.  Notice the relatively large variation
of L$_{FIR}$/L$_{opt}$ between the different galaxies.\\
\label{fig.sed}
}

At $z=2.515$ the spatial resolution of the {\it HST} images, corresponds
to $\sim 0.1$\,kpc, taking into account the gravitational
magnification.  Thus the colour gradient within the background galaxy
is on 1--2\,kpc scales -- similar to the obscured region in nuclear
starbursts in ultraluminous infrared galaxies at the present-day. A
constraint on the size of the submm emission region is obtained by subtracting
point sources from the 850 and 450\,$\mu$m maps. This procedure leaves
no detectable residual of extended emission, indicating that to our
measurement accuracy we are dealing with point sources. The tightest
constraint is obtained at 450\,$\mu$m where we find that the intrinsic
extent of the emission is less than $4''$, which corresponds to $\sim
8$\,kpc. There is thus no indication for a highly extended obscured
starburst in this relatively low-luminosity submm galaxy.

The precise redshift we have measured for SMM\,J16359+6612 enables us
to identify that this galaxy lies at an identical redshift to that of
\#273 and the \#384/\#468 multiple image system.  Using the lens model 
we estimate that SMM\,J16359+6612 lies between the \#384/\#468 and \#273 
galaxies in the source plane, and all three galaxies are less than 130\,kpc
apart. If these galaxies were not magnified by the cluster lens, all
three galaxies would appear within a radius of 8$''$.  Moreover, we can
place a limit of $\ls 100$\,km\,s$^{-1}$ on the possible velocity
offset between these three galaxies. They are thus all part of a single
group and it is likely that they are interacting, which may explain the
activity we detect in the submm waveband.  The strong clustering of the
submm galaxies with UV-bright populations highlights the opportunity
for measuring the distances to submm galaxies from the redshifts of
less-obscured companions, as well as the possible confusion which may
arise when trying to relate UV- and submm-selected populations, in the
absence of precise positions from radio counterparts.

SMM\,J16359+6612 has much redder colour in $(V-I)$ and $(I-K)$ than the
two other nearby UV-selected galaxies, confirming the dustier nature of
this galaxy.  If we compare the unlensed star formation rate (SFR) for
this three systems, we find based on their UV continuum that:
SMM\,J16359+6612 has 6 M$_\odot$\,yr$^{-1}$ (although certainly
underestimated due to dust extinction) \#468: 14 M$_\odot$\,yr$^{-1}$
and \#273: 4 M$_\odot$\,yr$^{-1}$.  For SMM\,J16359+6612 we can compare
the three estimates of its amplification-corrected star formation rate:
6 M$_\odot$\,yr$^{-1}$ from its dust-corrected UV luminosity,
11\,M$_\odot$\,yr$^{-1}$ from the H$\alpha$ flux (uncorrected for
extinction or aperture losses) and 500\,M$_\odot$\,yr$^{-1}$ based on
the far-infrared emission.  Applying the median extinction correction
derived for mid-IR selected luminous infrared galaxies at $z\sim 0.7$
by Flores et al.\ (2003), A$_{{\rm H}\alpha}\sim 2.1$, and a modest
correction for slit losses would increase the H$\alpha$-derived star
formation rate by a factor of $\sim 10\times$.  These results suggest
the vast majority of the young stars in this galaxy are obscured by
dust and are undetectable in the restframe far-UV.  Although it is
10$\times$ less luminous than the typical submm-selected galaxy studied
in blank-field surveys, this system shares the same extreme levels of
obscuration seen in the more luminous galaxies, rather than the
more modest dust obscuration inferred for the somewhat less
luminous UV-selected galaxies at these redshifts. We also note that the
H$\alpha$ line width and velocity structure, if they reflect the
dynamics of the galaxy, suggest that this system is more massive for
its UV luminosity than typical UV-selected galaxies at this epoch (Erb
et al.\ 2003).
 
In summary, we have identified a multiply-imaged submm galaxy seen
through the core of the rich cluster A\,2218.  The cluster lens
amplifies the background galaxy by a factor of $\sim 45\times$, providing a
high signal-to-noise view of an example of the sub-mJy submm population
which provides the bulk of the extragalactic background in this
waveband.  We estimate a redshift for this galaxy from our
highly-constrained lens models for the cluster and confirm this using
optical and near-infrared spectroscopy from Keck.  The redshift of the
submm galaxy is $z=2.516$, placing it at the same redshift as two other
strongly-lensed UV-bright galaxies in this field.  Our Keck
spectroscopy suggests that the emission from the submm galaxy is
dominated by star formation and in contrast to the typical UV spectral
properties of more luminous submm galaxies, this galaxy shows
Ly-$\alpha$ absorption, rather than emission (c.f.\ Chapman et al.\
2003).  The star formation rates we derive from the UV continuum,
H$\alpha$ and far-infared emission show that most of the star formation
activity is obscured and we suggest it is likely to be located in
the component $\gamma$, given its extreme $(I-K)$ colour.  
Because of its faint submm flux this galaxy is likely to be a good
example of the type of galaxy that makes most contribution to
the star formation history, thus deserving a detailled study with submm/mm 
interferometer to study the dynamics and mass of this obscured region.
The presence of three highly-amplified $z=2.515$ galaxies in our survey
field indicates that the submm galaxy resides in a compact group,
interactions within which may help explaining the triggering of the
obscured starburst we detect.

\section*{Acknowledgements}

We are grateful to R.\ Blandford, A.\ Fruchter, R.\ Hook,
C.\ Packham, J.\ Peacock, N.\ Reddy, G.\ Smith, R.\ Tilanus 
and T.\ Treu for their
respective roles in the acquisition and reduction steps for the data
that this paper is based upon.  We thank the anonymous referee for 
his constructive and useful report.
JPK acknowledges support from CNRS and Caltech. 
IRS acknowledges support from the Royal Society and the
Leverhulme Trust. AWB acknowledges support from the NSF under grant
number AST-0205937. We also acknowledge support from the UK--French
ALLIANCE collaboration programme \#00161XM.  The JCMT is operated by
the Joint Astronomy Centre on behalf of the United Kingdom Particle
Physics and Astronomy Research Council (PPARC), the Netherlands
Organization for Scientific Research, and the National Research Council
of Canada. Some of this data was obtained at the W.~M.\ Keck Observatory,
operated as a scientific partnership among Caltech, the University of
California and NASA. The Observatory was made possible by the generous
financial support of the W.~M.\ Keck Foundation. The William Herschel
Telescope is operated on the island of La Palma by the Isaac Newton
Group in the Spanish Observatorio del Roque de los Muchachos of the
Instituto de Astrofisica de Canarias.  This paper is based on
observations obtained with the NASA/ESA {\it Hubble Space Telescope},
obtained from the Data Archive of the Space Telescope Science Institute
which is operated by the Association of Universities for Research in
Astronomy, Inc., under NASA contract NAS5-26555.

\end{document}